\documentclass[a4paper,11pt]{article}
\usepackage{pos}
\usepackage{graphicx}
\usepackage{amsmath}

\usepackage{soul}

\title{Charmed Baryon Flow in the Hybrid Model}
\author[a]{Andrea Beraudo}
\affiliation[a]{INFN --- Sezione di Torino, Via P.~Giuria 1, I-10125 Torino, Italy}

\author*[b]{Jean F. Du Plessis}
\affiliation[b]{Center for Theoretical Physics -- a Leinweber Institute, Massachusetts Institute of Technology, Cambridge, MA 02139}

\author[c,d]{Daniel Pablos}
\affiliation[c]{Departamento de F\'isica, Universidad de Oviedo, Avda. Federico Garc\'ia Lorca 18, 33007 Oviedo, Spain}
\affiliation[d]{Instituto Universitario de Ciencias y Tecnolog\'ias Espaciales de Asturias (ICTEA), Calle de la Independencia 13, 33004 Oviedo, Spain}

\author[b]{Krishna Rajagopal}

\renewcommand{\hookBeforeTitle}{%
    \noindent\hfill{\small MIT-CTP/6152}\par
    \vspace{2em}
}

\emailAdd{beraudo@to.infn.it}
\emailAdd{jeandp@mit.edu}
\emailAdd{pablosdaniel@uniovi.es}
\emailAdd{krishna@mit.edu}
\abstract{Heavy quarks produced in heavy ion collisions can start out ultrarelativistic, lose energy, slow down, and ultimately diffuse in the strongly coupled quark-gluon plasma. We describe an extension of the Hybrid Strong/Weak Coupling Model that incorporates these regimes within the same framework that describes jet quenching. A composite prescription for energy loss, together with momentum diffusion and recombination with light partons from the medium, yields a reasonable description of heavy-flavor suppression and azimuthal anisotropies over a broad range of transverse momenta. The treatment of momentum diffusion in this implementation assumes Gaussian fluctuations. To assess the impact of recombination, we investigate the azimuthal anisotropy of $\Lambda_c$ baryons and confront our results with data from ALICE and CMS.}
\FullConference{13th International Conference on Hard and Electromagnetic Probes of High-Energy Nuclear Collisions (Hard Probes 2026)\\
21--26 June 2026\\
Vanderbilt University, Nashville, Tennessee, USA}

\newcommand{\kHQ}{\kappa_{\rm HQ}}
\newcommand{\ksc}{\kappa_{\rm sc}}
\newcommand{\RAA}{R_{AA}}
\newcommand{\pT}{p_T}

\begin{document}
\maketitle

\noindent
{\bf Introduction.}
Ultrarelativistic heavy ion collisions produce quark-gluon plasma (QGP), a strongly coupled deconfined thermal state of quarks and gluons, thought to have filled the universe shortly after the big bang~\cite{Busza:2018rrf}. However, studying this primordial fluid is no easy task. Its short lifetime precludes the use of external probes, which one would use to study more tame states of matter. We are therefore limited to using only probes formed during the collision itself to study the properties of this extreme state of matter.  The so-called hard probes have energy scales much larger than that of the plasma. This separation of scales helps us to calculate how the medium modifies the probes, making it possible to infer properties of the medium from measurements of the probes.
Two prototypical examples of hard probes~\cite{Apolinario:2022vzg} are jets, collimated showers of partons, and heavy quarks, which will be our focus in this work. 
Heavy-quark masses are much larger than the QGP temperature, which suppresses their thermal production, meaning that observed heavy hadrons come from heavy quarks produced in the initial collision that have propagated through the droplet of QGP as it expands and cools. 
In order to interpret the modification of the momentum distribution of heavy hadrons in PbPb collisions relative to that in pp collisions, however, one also needs to understand how the medium modifies heavy-quark hadronization~\cite{Altmann:2024kwx}. 
Several frameworks for studying the recombination of heavy quarks with partons from the medium exist in the literature~\cite{Plumari:2017ntm,He:2019vgs,Beraudo:2022dpz}. In this work, we
incorporate the local color neutralization (LCN) treatment of recombination~\cite{Beraudo:2022dpz} into the Hybrid Model --- originally developed to describe jets in heavy ion collisions --- allowing us to investigate how the interplay between 
heavy quark energy loss, diffusion, and recombination
affects the flow of charmed baryons and mesons. 

\noindent
{\bf Heavy quarks in the Hybrid Model.}
Jets and heavy quarks are often considered separately as probes of QGP. We wish to describe both within a single framework, including heavy quarks that are initially relativistic and subsequently slow down and diffuse in the medium. In Ref.~\cite{Beraudo:2025nvq}, we extended the Hybrid Strong/Weak Coupling Model~\cite{Casalderrey-Solana:2014bpa,Casalderrey-Solana:2016jvj,
Casalderrey-Solana:2018wrw} to do exactly that.

The Hybrid Model combines perturbative parton showers generated by \textsc{Pythia}~8~\cite{Bierlich:2022pfr} with a strongly coupled description of energy loss 
in a hydrodynamic medium~\cite{Shen:2014vra}. 
Holographic calculations in $\mathcal N=4$ supersymmetric Yang--Mills (SYM) theory provide descriptions of energy loss in the massless~\cite{Chesler:2014jva,Chesler:2015nqz} and infinite-mass~\cite{Herzog:2006gh,Gubser:2006bz} limits. For phenomenological purposes, we combine them in the prescription~\cite{Beraudo:2025nvq,DuPlessis:2025ntp}
\begin{equation}
 \begin{aligned}
 \frac{dE}{dx}&=-\min\!\left[
 \frac{4x^2E_{\rm in}}{\pi x_{\rm stop}^2\sqrt{x_{\rm stop}^2-x^2}},
 \eta_D\sqrt{E^2-M^2}\right],
 \end{aligned}
 \label{eq:composite}
\end{equation}
dubbed \textit{composite energy loss}.
Here $E_{\rm in}$ is the initial energy, $x$ is the distance traversed in the fluid rest frame, $M$ is the quark mass, and $T$ is the local temperature. We have also introduced the strongly coupled stopping distance $x_{\rm stop}\equiv(1/2\ksc)\left(E_{\rm in}/T^4\right)^{1/3}$ and drag coefficient $\eta_D\equiv\kHQ T^2/2M$. The relativistic heavy quark initially loses energy as a light quark does~\cite{Chesler:2014jva,Chesler:2015nqz}
but, as it slows, at a certain point
it switches to losing energy as a heavy quark does via drag~\cite{Herzog:2006gh,Gubser:2006bz}, 
with the switching point such that both the energy and the energy-loss rate remain continuous as functions of distance. The model parameters $\ksc$ and $\kHQ$ absorb differences between the SYM and QCD plasmas; we use $\ksc=0.404$ from a fit to light-hadron and jet suppression~\cite{Casalderrey-Solana:2018wrw} and $\kHQ=4.4$ from Ref.~\cite{Beraudo:2025nvq}.

\begin{figure}[t]
    \centering
    \includegraphics[width=0.49\linewidth]{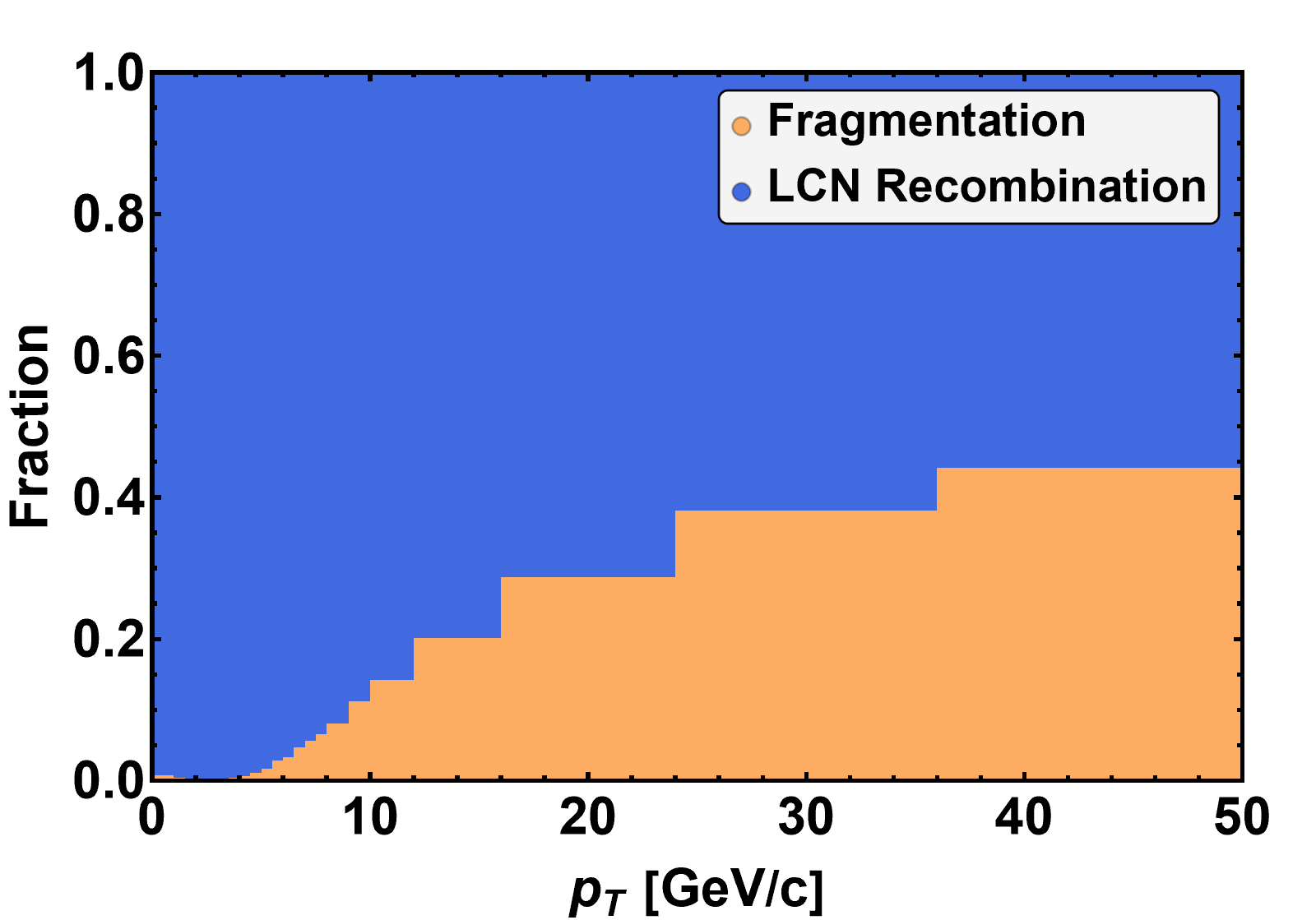}
    \includegraphics[width=0.49\linewidth]{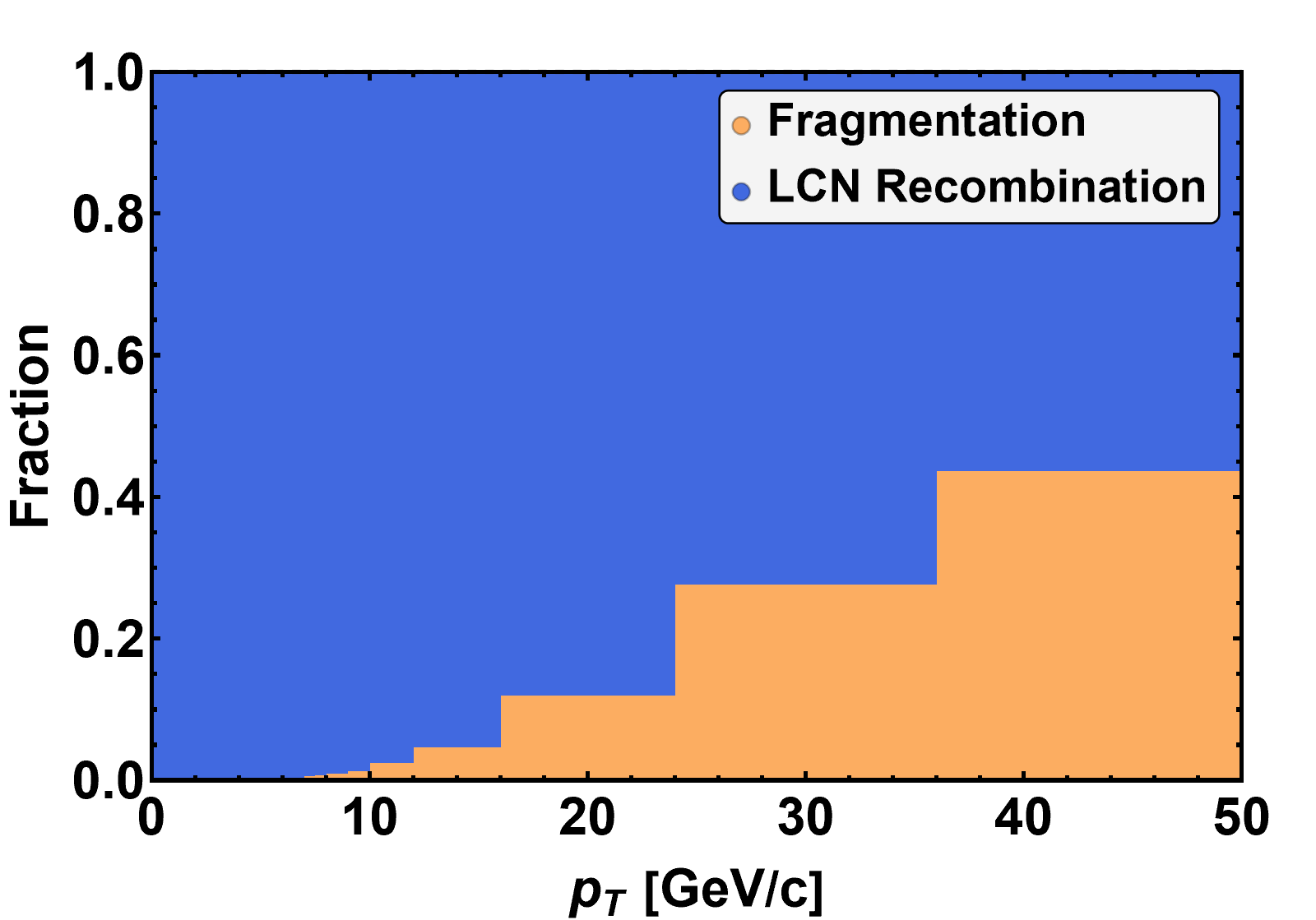}
    \caption{Fraction of $D^0$ mesons (left) and $\Lambda_c$ baryons (right) with a given $p_T$ in PbPb collisions with energy $\sqrt{s_{\rm NN}}=5.02$~TeV in the 0-5\% centrality class that are formed either via LCN recombination (blue) or via fragmentation as described by \textsc{Pythia}~8 (orange).  Almost every $D^0$ meson with $p_T$ less than about twice its mass and almost every $\Lambda_c$ baryon with $p_T$ less than about four times its mass (which is to say most $D^0$ mesons and almost all $\Lambda_c$ baryons) form via recombination. Figs.~from Ref.~\cite{Beraudo:2025nvq}.}
    \label{fig:hadform}
    \vspace{-0.05in}
\end{figure}

In the low-velocity regime where Eq.~\eqref{eq:composite} prescribes that heavy quarks lose energy via drag, we also add Gaussian momentum kicks. For simplicity, we choose  Gaussian fluctuations that are isotropic in the fluid rest frame, with equal longitudinal and transverse diffusion coefficients $\kappa_T=\kappa_L=\kHQ\gamma T^3$ satisfying the Einstein Relation\footnote{For Langevin dynamics, the Einstein relation is crucial for consistent equilibration. However, in a strongly coupled plasma this relation is badly violated
at relativistic velocities~\cite{Gubser:2006nz}, prohibiting the consistent simultaneous use of the strongly coupled drag and diffusion coefficients in a Langevin equation for relativistic heavy quarks.
The key to resolving this puzzle is that non-Gaussian momentum fluctuations become increasingly important at increasingly relativistic velocities~\cite{Rajagopal:2025ukd,Rajagopal:2025rxr,DuPlessis:2026pyr,Rajagopal:2026urq,Scheihing-Hitschfeld:2026lbg,DuPlessis:2026qvj}. However, in employing our composite energy loss ansatz \eqref{eq:composite}, we incorporate momentum fluctuations only in the heavy-quark drag regime at lower velocities, where deviations from the Einstein relation should be small. Preliminary studies incorporating non-Gaussian fluctuations~\cite{Rajagopal:2026urq,DuPlessis:2026qvj} suggest that their effects are indeed small at low velocity but become increasingly important at increasingly relativistic velocities.
}
$\kappa_L=2TE\eta_D$, and hence specified by the parameter $\kHQ$.
We reweight the initial charm spectrum using fixed-order next-to-leading-logarithm calculations~\cite{Cacciari:1998it,Cacciari:2012ny} and implement charm hadronization via a local color neutralization (LCN) model~\cite{Beraudo:2022dpz}. A charm quark can recombine with a thermal light antiquark or diquark, which transfers
momentum from the flowing medium to the resulting hadron~\cite{Lin:2003jy,Greco:2003vf}. 
However, this model for recombination is intended for heavy quarks at low to intermediate momenta, and one expects that at large transverse momentum the heavy quark should undergo vacuum-like fragmentation.
We describe our procedure for modeling this in detail in Ref.~\cite{Beraudo:2025nvq}. This procedure allows heavy quarks in a given event to either undergo recombination or vacuum-like fragmentation. In a given $p_T$ bin, some fraction of a given heavy hadron species will therefore have been formed by recombination, with the remaining fraction having been formed through vacuum-like fragmentation. Fig.~\ref{fig:hadform} shows the differences between the $p_T$-dependence of these  fractions for $D^0$ mesons and $\Lambda_c$ baryons.

\begin{figure}[ht]
 \centering
 \includegraphics[width=\linewidth]{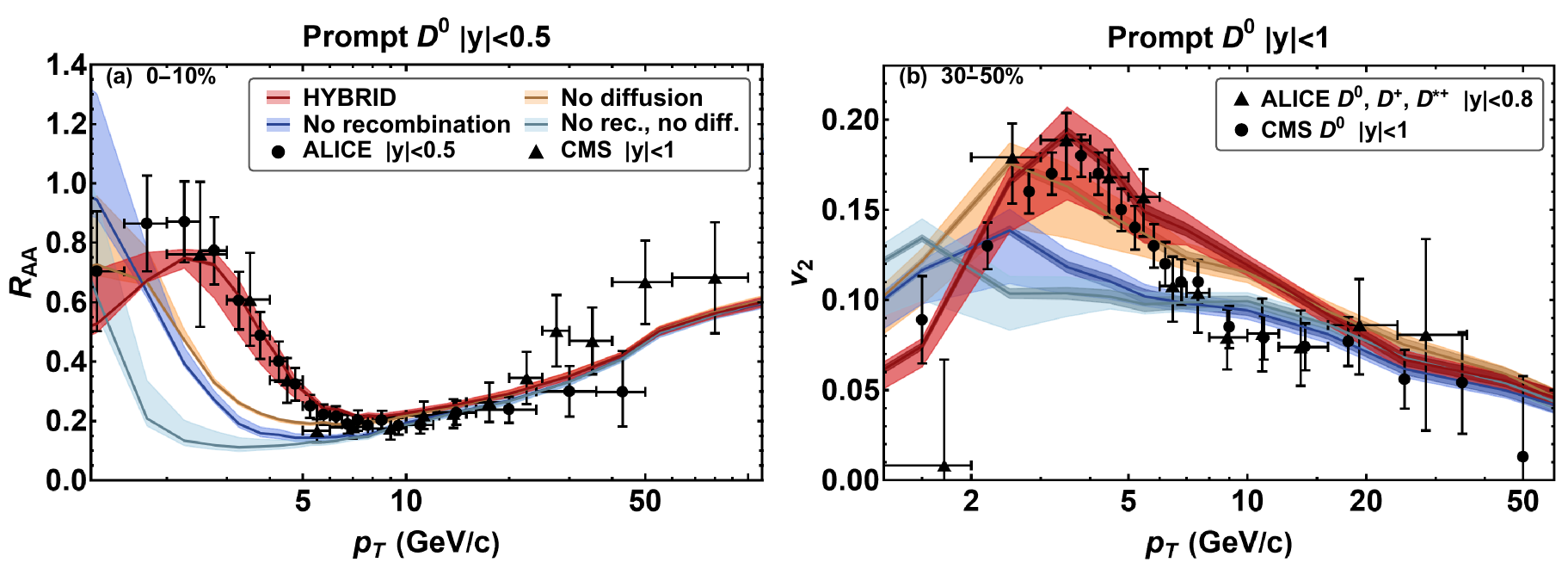}
 \caption{Prompt $D^0$ suppression (left) and elliptic flow (right) in Pb--Pb collisions at $\sqrt{s_{NN}}=5.02$~TeV, with $\kHQ=4.4$.  Results from the full Hybrid Model calculation are compared to results obtained without diffusion or recombination or without either. Narrow dark (wider light) bands indicate statistical uncertainties (uncertainties in the reweighting of the initial heavy quark spectrum).  Data are from ALICE~\cite{ALICE:2020iug,ALICE:2021rxa} and CMS~\cite{CMS:2017qjw,CMS:2020bnz}. Kinematic cuts in the  calculations in the left (right) panel are those for the ALICE (CMS) measurements shown as circles, with CMS (ALICE) measurements shown as triangles. Figs.~from Ref.~\cite{Beraudo:2025nvq}.}
 \label{fig:phenomenology}
 \vspace{-0.05in}
\end{figure}

Fig.~\ref{fig:phenomenology} illustrates the importance of including both diffusion and recombination effects to reproduce the behavior seen in the data. With $\kHQ=4.4$, we obtain a reasonable description of both the low-$\pT$ enhancement in prompt $D^0$ $\RAA$ and its elliptic flow. The same model describes $b$-jet suppression and a range of heavy-hadron observables~\cite{Beraudo:2025nvq}. 
It is pleasing to see both jets and heavy flavor observables across two orders of magnitude in transverse momentum described seamlessly and successfully within the same model framework.

This simultaneous description of a range of heavy quark meson data is very encouraging, but to have confidence in the control over the recombination dynamics, one also needs to be able to describe heavy quark baryons such as the $\Lambda_c$ simultaneously. We therefore confront our model with experimental measurements of the elliptic flow of $\Lambda_c$ baryons as well as $D^0$ mesons below.

\noindent
{\bf $\Lambda_c$ elliptic flow.}
A $\Lambda_c$ baryon that forms via LCN recombination between a charm quark and a diquark from the flowing QGP picks up a larger flow contribution to its momentum from the medium than a $D^0$ meson formed via recombination with a single (anti)quark does. Understanding the differences between the flow of $\Lambda_c$ baryons and prompt $D^0$ mesons therefore gives unique insights into the physics of recombination and how recombination transitions into the vacuum-like fragmentation expected at large transverse momenta. At equal transverse momenta of the hadrons, one also expects to be probing different kinematic regions of the charm transport, due to the larger mass and larger thermal contribution to charmed baryons versus mesons. ALICE~\cite{ALICE:2026zcz} has recently provided the first measurement of the elliptic flow of $\Lambda_c$ baryons, with their
data providing 3.7$\sigma$ evidence that they have 
a larger elliptic flow than $D^0$ mesons do at intermediate transverse momenta, in qualitative agreement with the expectation from LCN recombination. Recent preliminary CMS measurements~\cite{CMS:2026rmi} 
are statistically compatible with those
from ALICE, but show no evidence for a difference in the elliptic flow of $\Lambda_c$
baryons and $D^0$ mesons. This potential tension motivates
looking at what the Hybrid Model predicts.

\begin{figure}
    \centering
    \includegraphics[width=0.49\linewidth]{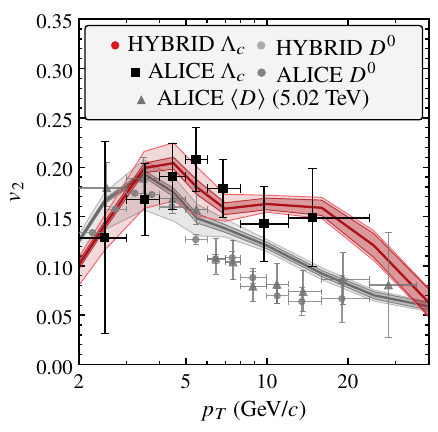}
    \includegraphics[width=0.49\linewidth]{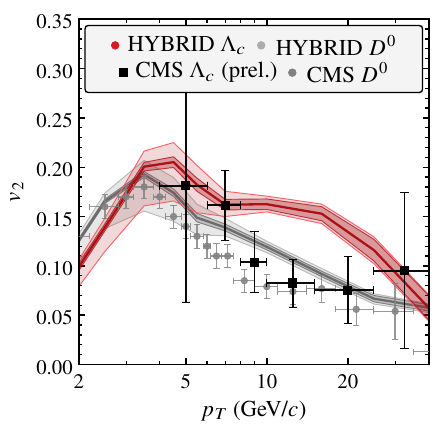}
    \caption{Hybrid Model results, from calculations done with $\kHQ=4.4$, for the elliptic flow of prompt $\Lambda_c$ baryons (red) and $D^0$ mesons (grey) in 30--50\% central Pb--Pb collisions with $\sqrt{s_{\rm NN}}=5.02$~TeV with $|y|<0.8$ (left) and $|y|<1$ (right) compared with ALICE (left)~\cite{ALICE:2026zcz} and CMS (right)~\cite{CMS:2026rmi,CMS:2020bnz} measurements. Black squares denote $\Lambda_c$ measurements and grey circles $D^0$ measurements. Grey triangles show the older ALICE average of prompt $D^0$, $D^+$ and $D^{*+}$ flow~\cite{ALICE:2020iug}. The newer ALICE measurements and CMS $\Lambda_c$ data are at $\sqrt{s_{\rm NN}}=5.36$~TeV; the older ALICE average and CMS $D^0$ data are at $\sqrt{s_{\rm NN}}=5.02$~TeV. The CMS $\Lambda_c$ 
    data~\cite{CMS:2026rmi} are preliminary. Dark (light) model bands show statistical uncertainties (uncertainties in the reweighting of the initial heavy quark spectrum). Experimental error bars include statistical and systematic uncertainties combined in quadrature.}
    \label{fig:lambdac}
    \vspace{-0.05in}
\end{figure}

The Hybrid Model calculations shown in 
Fig.~\ref{fig:lambdac} confirm the qualitative expectation from the physics of LCN 
recombination. As in the ALICE data, there is a clear separation between the $v_2$ of $\Lambda_c$ baryons and $D^0$ mesons.
In fact, the Hybrid Model results for $\Lambda_c$ $v_2$ are in better agreement with ALICE data than is the case for $D^0$ $v_2$. There is, however, some tension between Hybrid Model calculations and CMS measurements of $\Lambda_c$ $v_2$. As the experimental uncertainties are reduced with future high statistics measurements, this promises to be a new way to constrain the dynamics of recombination and its interplay with vacuum-like fragmentation.

Since comparing the elliptic flow of the baryons vs mesons at equal transverse momentum probes charm quarks at different momenta, a good treatment of charm transport is important. Future comparisons between phenomenological models will need to carefully disentangle differences in the treatment of transport from differences in the treatment of hadronization~\cite{Rapp:2018qla,Zhao:2023nrz}.

\noindent
{\bf Outlook.}
Baryon-to-meson ratios~\cite{ALICE:2021bib} and the elliptic flow of charmed baryons and mesons provide complementary tests of in-medium hadronization. The increased statistics from the full Run 3 dataset as well as future heavy ion runs promise to put increasingly stringent constraints on models. It is therefore important to build a more and more robust understanding of what features of the data should be attributed to hadronization vs transport. The motivating goal of including heavy quarks in the Hybrid Model is to obtain a seamless description of heavy quark suppression and flow over a large range of momenta in the same model framework in which the modification to the substructure of jets --- including heavy quark jets --- can be described. This promises to provide a whole new suite of observables with which to disentangle and constrain heavy quark transport and hadronization, parton energy loss, and the response of the medium to jets and heavy quarks.

\noindent 
{\bf Acknowledgments.} Research supported by the U.S.~DOE, Office of Science, Office of Nuclear Physics under grant Contract Number DE-SC0011090, by the Ram\'on y Cajal fellowship RYC2023-044989-I and
by MUR within the Prin 2022sm5yas project.
This work made use of SubMIT~\cite{Bendavid:2025arv}.

\vfill

\bibliographystyle{JHEP}
\bibliography{proceedings}
\end{document}